\documentclass{article}
\usepackage[left=1.5in,right=1.5in,top=1.5in,bottom=1.5in]{geometry}

\usepackage{booktabs} 

\usepackage{graphicx}
\usepackage[caption=false,font=footnotesize]{subfig}
\usepackage{multirow}
\usepackage{amsmath}
\usepackage{authblk}

\DeclareMathOperator*{\argmin}{arg\,min}

\begin{document}

\title{Accelerating Large-Scale Data Analysis by Offloading to High-Performance Computing Libraries using Alchemist\footnote{Accepted for publication in Proceedings of the 24th ACM SIGKDD International Conference on Knowledge Discovery and Data Mining, London, UK, 2018}}

\author[1]{Alex Gittens\thanks{{\tt gittea@rpi.edu}}}
\author[2]{Kai Rothauge\thanks{{\tt kai.rothauge@berkeley.edu}}}
\author[2]{Shusen Wang}
\author[2]{Michael W. Mahoney}
\author[3]{Lisa Gerhardt}
\author[3]{Prabhat}
\author[2]{Jey Kottalam}
\author[4]{Michael Ringenburg}
\author[4]{Kristyn Maschhoff}
\affil[1]{Rensselaer Polytechnic Institute, Troy, NY}
\affil[2]{UC Berkeley, Berkeley, CA}
\affil[3]{NERSC/LBNL, Berkeley, CA}
\affil[4]{Cray Inc., Seattle, WA}

\maketitle

\begin{abstract}
Apache Spark is a popular system aimed at the analysis of large data sets, but recent studies have shown that certain computations---in particular, many linear algebra computations that are the basis for solving common machine learning problems---are significantly slower in Spark than when done using libraries written in a high-performance computing framework such as the Message-Passing Interface (MPI).

To remedy this, we introduce Alchemist, a system designed to call MPI-based libraries from Apache Spark. Using Alchemist with Spark helps accelerate linear algebra, machine learning, and related computations, while still retaining the benefits of working within the Spark environment. We discuss the motivation behind the development of Alchemist, and we provide a brief overview of its design and implementation. 

We also compare the performances of pure Spark implementations with those of Spark implementations that leverage MPI-based codes via Alchemist. To do so, we use data science case studies: a large-scale application of the conjugate gradient method to solve very large linear systems arising in a speech classification problem, where we see an improvement of an order of magnitude; and the truncated singular value decomposition (SVD) of a 400GB three-dimensional ocean temperature data set, where we see a speedup of up to 7.9x. We also illustrate that the truncated SVD computation is easily scalable to terabyte-sized data by applying it to data sets of sizes up to 17.6TB.
\end{abstract}

\section{Introduction}
\label{sec:introduction}


Apache Spark~\cite{Zaharia2016} is a cluster-computing framework developed for the processing and analysis of the huge amount of data being generated over a wide range of applications, and it has seen substantial progress in its development and adoption since its release in 2014. 
It introduced the resilient distributed dataset (RDD) \cite{Zaharia2012}, which can be cached in memory, thereby leading to significantly improved performance for iterative workloads often found in machine learning. 
Spark provides \emph{high productivity computing} interfaces to the data science community, and it includes extensive support for graph computations, machine learning algorithms, and SQL queries. 

An altogether different framework for distributed computing is provided by the message-passing parallel programming model, which mediates cooperative operations between processes, where data is moved from the address space of one process to that of another. 
The Message Passing Interface (MPI)~\cite{mpi2015} is a standardized and portable message-passing specification that is favored in the \emph{high performance computing} (HPC) community because of the high performance that can be attained using well-written MPI-based codes. 
Although originally developed for distributed memory systems, the modern MPI standard also supports shared memory and hybrid systems. 
Popular implementations of the MPI standard are MPICH and Open MPI, with numerous derivatives of MPICH having been developed over the years. 
While bindings for MPI are available for some popular languages such as Python, R, and Julia, the vast majority of MPI-based libraries are written in C, C++, or Fortran, languages that are generally less accessible to data scientists and machine learning practitioners. 
Also, MPI offers no fault tolerance, nor does it offer support for elasticity.

In this paper, we introduce Alchemist~\cite{alchemist2018}, a system developed to facilitate using existing MPI-based libraries from Apache Spark. The motivation for Alchemist is that, while Spark does well for certain types of data analysis, recent studies have clearly shown stark differences in runtimes when other computations are performed in MPI, compared to when they are performed in Spark, with the high-performance computing framework consistently beating Spark by an order of magnitude or more.  Some of the applications investigated in these case studies include distributed graph analytics~\cite{Slota2016}, and k-nearest neighbors and support vector machines~\cite{ReyesOrtiz2015}. However, it is our recent empirical evaluations~\cite{Gittens2016} that serve as the main motivation for the development of Alchemist. Our results in~\cite{Gittens2016} illustrate the difference in computing times when performing certain matrix factorizations in Apache Spark, compared to using MPI-based routines written in C or C++ (C+MPI).

The results we showed for the singular value decomposition (SVD), a very common matrix factorization used in many different fields, are of particular interest. The SVD is closely related to principal component analysis (PCA), a method commonly used to discover low-rank structures in large data sets. 
SVD and PCA are particularly challenging problems for Spark because the iterative nature of their algorithms incurs a substantial communication overhead under the bulk synchronous programming model of Spark. 
The upshot is that not only is a Spark-based SVD or PCA more than an order of magnitude slower than the equivalent procedure implemented using C+MPI for data sets in the 10TB size range, but also that Spark's overheads in fact dominate and its performance anti-scales\footnote{As the number of nodes increases, the overheads take up an increasing amount of the computational workload relative to the actual matrix factorization calculations.}. 
(See Figures 5 and 6 in~\cite{Gittens2016} for details.)
There are several factors that lead to Spark's comparatively poor performance, including scheduler delays, task start delays, and task overheads~\cite{Gittens2016}.   

Given this situation, it is desirable to call high-performance libraries written in C+MPI from Spark to circumvent these delays and overheads.
Done properly, this could provide users with the ``best of both worlds.'' 
That is, the MPI-based codes would provide \emph{high performance}, e.g., in terms of raw runtime, while the Spark environment would provide \emph{high productivity}, e.g., by providing convenient features such as fault-tolerance by allowing RDDs to be regenerated if compute nodes are lost, as well as a rich ecosystem with many data analysis tools that are accessible through high-level languages such as Scala, Python, and R. 

We envision that Alchemist will be used as part of a sequence of data analysis operations in a Spark application. Some of these operations may be handled satisfactorily by Spark itself, but when MPI-based libraries are available that could perform critical computations faster, then the user can easily choose to call these libraries using Alchemist instead of using Spark's implementations. This allows us to leverage the efforts of the HPC community to bypass Spark's high overheads where possible. 

To accomplish this, Alchemist spawns MPI processes when it starts, and then it dynamically links to a given MPI-based library as required. The data from an RDD is transferred from the Spark processes to the MPI processes using sockets, which we have found to be an efficient method to handle the transfer of large distributed data sets between the two frameworks.

To assess and illustrate the improvement in performance that can be achieved by using Alchemist, we describe a comparison of the timing results for two important linear algebra computations.
\begin{itemize}
\item \textit{Conjugate gradient (CG)}. The CG method is a popular iterative method for the solution of linear systems of equations where the underlying matrix is symmetric positive-definite. We compare the performance of a custom implementation of CG in Spark to that of a modified version of an MPI-based implementation found in the Skylark library~\cite{Skylark2016} (a library of randomized linear algebra~\cite{Drineas2016} routines).
\item \textit{Truncated SVD}. The truncated SVD is available in MLlib, Spark's library for machine learning and linear algebra operations. We compare the performance of Spark's implementation with that of a custom-written MPI-based implementation (that also uses methods from randomized linear algebra~\cite{Drineas2016}) that is called via Alchemist.
\end{itemize}
Both of these computations are iterative and therefore incur significant overheads in Spark, which we bypass by instead using efficient C+MPI implementations. This, of course, requires that the distributed data sets be transmitted between Spark and Alchemist. As we will see, this does cause some non-negligible overhead; but, even with this overhead, Alchemist can still outperform Spark by an order of magnitude or more.\footnote{Alchemist is so-named since it solves this ``conversion problem'' (from Spark to MPI and back) and since the initial version of it uses the Elemental package~\cite{elemental2017}.}

The rest of this paper is organized as follows. Combining Spark with MPI-based routines to accelerate certain computations is not new, and in Section 2, we discuss this related work. 
In Section 3, we give an overview of the design and implementation of Alchemist, including how the MPI processes get started, how data is transmitted between Spark and Alchemist, Alchemist's use of the Elemental library, and a brief discussion of how to use Alchemist. 
In Section 4, we describe the experiments we used to compare the performance between Spark and Alchemist, including our SVD computations on data sets of sizes up to 17.6TB. We summarize our results and discuss some limitations and future work in Section 5.

Note that, in this paper that introduces Alchemist, our goal is simply to provide a basic description of the system and to illustrate empirically that Spark can be combined with Alchemist to enable scalable and efficient linear algebra operations on large data sets. 
A more extensive description of the design and usage of Alchemist can be found in a companion paper~\cite{Gittens2018b}, as well as in the online documentation~\cite{alchemist2018}.


\section{Related Work}
\label{sec:related_work}

Several recent projects have attempted to interface Spark with MPI-based codes. One of these is Spark+MPI~\cite{Anderson2017}, which also invokes existing MPI-based libraries. The approach used by this project serializes the data and transfers it from Spark to an existing MPI-based library using shared memory. The output from the computations is then written to disk using the fault-tolerant HDFS format, which can then be read by Spark and loaded into an RDD. Spark+MPI offers a simple API and supports sparse data formats, but the reported overheads of this system are significant when compared to the compute time, and the data sets used were relatively small, so it is not yet clear if this approach is scalable or appropriate for large, dense data sets.

Another approach is taken by Spark-MPI~\cite{Malitsky2017}, which supplements Spark's driver-worker model with a Process Management Interface (PMI) server for establishing MPI inter-worker communications. While promising, the project is still in its early stages, and it is unclear if or how it is possible to call MPI-based libraries from Spark.

Smart-MLlib~\cite{Siegal2016} allows Spark to call custom machine-learning codes implemented in C++ and MPI, with data transfer between Spark's RDD and the MPI-based codes accomplished by writing and reading from files. The system has not been shown to be scalable and appears to no longer be under active development.

\section{Overview of Alchemist}
\label{sec:overview}

In this section, we give a high-level overview of Alchemist. The design and implementation of the Alchemist framework is discussed in Section~\ref{subsec:design}, and how data is transmitted between Spark and Alchemist, and then stored in Alchemist, is addressed in Section~\ref{subsec:transmitting_data}. A quick note on the required dependencies and an introduction to the API is given in Section~\ref{subsec:usage}.

We refer the interested reader to our companion paper~\cite{Gittens2018b} and the online documentation~\cite{alchemist2018} for a more extensive discussion of the design, implementation, and usage aspects of Alchemist.

\subsection{Design and Implementation}
\label{subsec:design}

\begin{figure}[htp]
\centering
\includegraphics[width=\textwidth]{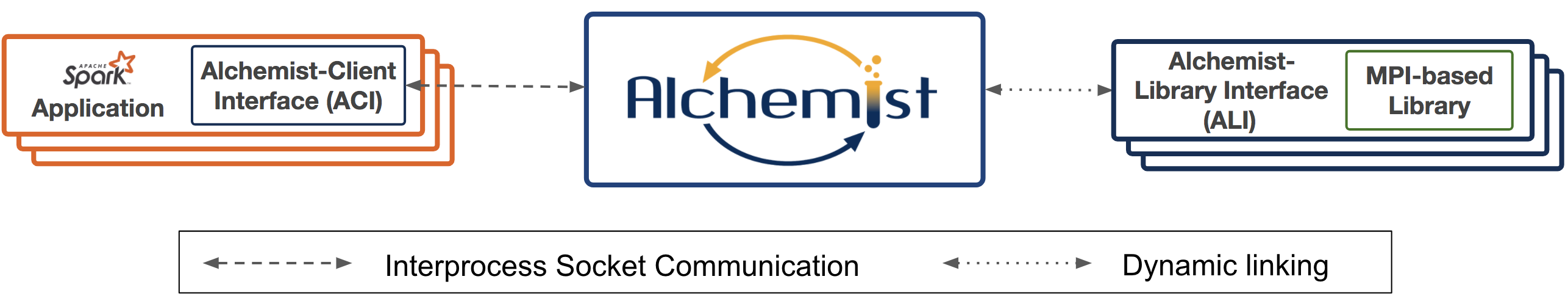}
\caption{Outline of the Alchemist framework. One or more Spark applications can each call one or more MPI-based libraries to help accelerate some numerical computations. Alchemist acts as an interface between the applications and the libraries, with data between Alchemist and the applications transmitted using TCP sockets. Alchemist loads the libraries dynamically at runtime.}
\label{fig:alchemist_framework}
\end{figure}

The Alchemist framework is shown in Figure~\ref{fig:alchemist_framework} and consists of three key components.

\subsubsection{\textbf{Alchemist}} 

The core Alchemist system acts as a bridge between an Apache Spark application and one or more MPI-based libraries. Alchemist has a driver process and several worker processes that communicate using MPI, and the MPI-based libraries run on these processes. Alchemist has a server-based architecture, with one or more Spark applications being able to connect to it concurrently, assuming sufficient workers are available. 

The current implementation requires Alchemist to run on a user-specified number of nodes, and the Spark application to run on a different set of nodes. The Spark application then connects to Alchemist and transfers the relevant data to it over the network. The application requires the Alchemist-Client Interface in order to connect to Alchemist, which we discuss next.

\subsubsection{The \textbf{Alchemist-Client Interface (ACI)}}

The ACI allows the Spark application to communicate with Alchemist. The main computational overhead of using Alchemist is the time it takes to transmit large data sets from the Spark application to Alchemist and back. Our approach to transferring the data from the Spark processes directly to the MPI processes is to use TCP sockets. This is an in-memory procedure and is therefore very fast, while not requiring an additional copy of the data set, and therefore the most practical option for our purposes. Other possibilities would be to transfer the data using files, or to use some in-memory intermediary in a format that both Spark and Alchemist can use. See Section~\ref{subsec:transmitting_data} for further discussion on this.

The ACI opens multiple TCP sockets between the Spark executors and Alchemist workers, and one socket connection between the two driver processes. The communication is asynchronous, allowing not only other Spark drivers to connect to the Alchemist driver concurrently, but also to accommodate the case where each Alchemist worker receives data from several Spark~executors. 

Non-distributed data, such as metadata and input arguments such as step sizes, maximum iteration counts, cut-off values, etc., are transferred between the Spark and Alchemist drivers. 

Distributed data, on the other hand, is transferred between the Spark executors and the Alchemist workers. When transferring the data from an RDD to a recipient Alchemist worker, the Spark executor sends each row of the RDD partitions to the recipient worker by transmitting the row as sequences of bytes. The received data is then recast to floating point numbers on the MPI side. Conversely, the transmission of matrices from MPI to Spark is facilitated by Alchemist in a similar row-wise fashion. In the applications we consider in this paper, the distributed data sets are dense matrices of floating point numbers. Concretely, this means that Alchemist currently sends and receives data using Spark's {\tt IndexedRowMatrix} RDD data structure.

\subsubsection{The \textbf{Alchemist-Library Interface (ALI)}}

To communicate with a library, Alchemist calls it through an ALI. The ALIs are shared objects written in C/C++ that import their associated MPI-based libraries and provide a generic interface through which Alchemist is able to access the MPI-based routines. Alchemist will load an ALI dynamically at runtime if it is required by some Spark application.

Alchemist sends the name of the routine in the library that the application wants to use, as well as the serialized input parameters, to the ALI. After deserializing the input parameters, the ALI calls the routine in the MPI-based library with the supplied input in a format appropriate for that specific library. Once the computations have completed, the results get serialized and sent to Alchemist, which then passes them on to the application over the network, where they are deserialized by the ACI.

Structuring the ALIs as shared object that get linked to at runtime keeps Alchemist flexible by avoiding the need to maintain a wrapper function inside Alchemist for every function in each MPI-based library that an application may want to call. Such a centralized system would incur a significant amount of work for each additional library that gets added, would impose a severe maintenance and support burden, and would not give users the option to add libraries in a portable manner.

\subsection{Transmitting and Storing Distributed Data}
\label{subsec:transmitting_data}

A key component of the Alchemist system is the efficient transmission of distributed data sets between the Spark application and Alchemist. In particular, in order for Alchemist to serve as a bridge, it must support sending distributed input data sets from the application to the library, and returning distributed output data sets (if any) from the library to the application. The design goals of the Alchemist system include making the transmission of these distributed data sets easy to use, efficient, and scalable.

Broadly, there are three approaches that could be used to transmit the data:
\begin{itemize}
\item File I/O: Writing the distributed data set to a distributed file format on one side and then reading it on the other has the benefit of being easy to use and, if HDFS is used, fault-tolerant. This approach will generally tend to be very slow when working with standard HDDs (although it has been argued that using an array of SSDs as the storage platform would alleviate this problem).
\item 
Intermediate Storage:
This approach would provide an intermediate form of storage of the data in memory that can be read by both sides. This could be done using shared memory, or in-memory distributed storage systems. Since we are considering very large data sets, having a third copy of the data in memory (on top of the {\tt IndexedRowMatrix} in the application and the {\tt DistMatrix} in Alchemist) is undesirable.
\item 
Direct Transfer:
Our preferred approach is to transfer the data from the Spark processes directly to the MPI processes using TCP sockets. This is much faster than writing to file and does not require an additional copy of the data set. 
\end{itemize}

As the Alchemist workers receive the data arriving through the sockets from the Spark executors, they store it in a distributed matrix using the \textit{Elemental}~\cite{elemental2017} library. Elemental is an open-source software package for distributed-memory dense and sparse-direct linear algebra and optimization. It provides a convenient interface for handling distributed matrices with its {\tt DistMatrix} class, which is what Alchemist uses to store the data being transmitted from the RDDs. Elemental also provides a large suite of sequential and distributed-memory linear algebra operations that can be used to easily manipulate the distributed matrices. Copying data from distributed data sets in Spark to distributed matrices in Elemental requires some changes in the layout of the data, a task that is handled by Alchemist. The Elemental distributed matrix then serves as input to the C+MPI routines in the MPI-based library that Alchemist~calls. 

\subsection{Using Alchemist}
\label{subsec:usage}

Alchemist is designed to be easily deployable, flexible, and easy to use. The only required imports in a Spark application are the ACI and, optionally, library wrappers for the MPI-based libraries that the Spark application needs to access. Alchemist works with standard installations of Spark.

\subsubsection{Dependencies}

To build the core Alchemist system, the user must have a common implementation of MPI 3.0 or higher~\cite{mpi2015} installed, such as recent versions of Open MPI~\cite{openmpi2018} or MPICH~\cite{mpich2018}. Alchemist also requires the \textit{Boost.Asio} library~\cite{boostasio2018} for communicating with the ACI over sockets, and the \textit{Elemental} library~\cite{elemental2017} discussed previously for storing the distributed data and providing some linear algebra operations. In addition, each MPI-based library (and its dependencies) that the Spark application wants to use should be installed on the system, since these are needed when compiling the ALI.

\subsubsection{The API}

\begin{figure}[htp]
\centering
\includegraphics[width=0.8\textwidth]{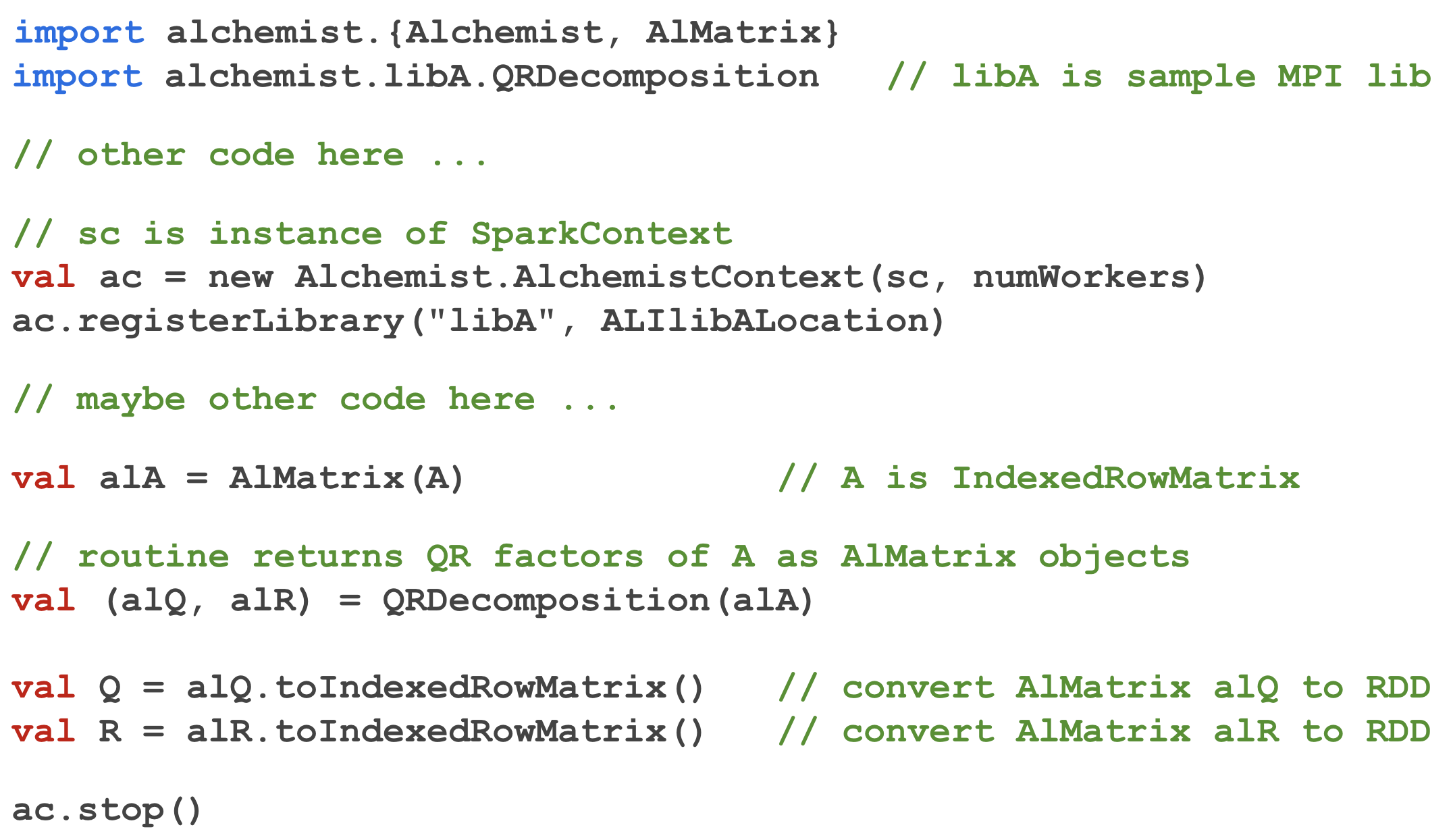}
\caption{An excerpt of Spark code that calls Alchemist to perform the QR decomposition of the {\tt IndexedRowMatrix} {\tt A} using the hypothetical MPI-based library libA.}
\label{fig:samplelisting}
\end{figure}

The API for Alchemist is restricted to the user's interaction with the ACI within the Spark application. Figure~\ref{fig:samplelisting} shows an excerpt that illustrates how to use Alchemist in a Spark application. In this sample, LibA is a hypothetical MPI-based library that implements a routine for performing the QR decomposition of a matrix. The Spark user wants to call this routine to find the decomposition of the {\tt IndexedRowMatrix} {\tt A}.

The example has the following components that we will briefly describe:
\begin{itemize}
\item The application imports the ACI (called {\tt Alchemist} here). It also imports the \textit{library wrapper} {\tt libA} for the MPI-based library it wants to work with, and the object that represents the function it wants to use ({\tt QRDecomposition} in this case). Library wrappers simplify the syntax for calling routines in the MPI-based libraries, and each MPI-based library can have its own library wrapper. Library wrappers are optional, but for brevity we do not go into more detail here. (See~\cite{Gittens2018b} for more details.)
\item The {\tt AlchemistContext} instance requires the {\tt SparkContext} instance {\tt sc} and the number of workers that the application wants to use.
\item Registering the library with the {\tt AlchemistContext}. Alchemist needs to know where to find the ALI for libA on the system, so the Spark application needs to register the library with Alchemist by giving it the library name and file path to where it is located.
\item Alchemist uses {\tt AlMatrix} objects, which are matrix handles that act as proxies for the distributed data sets stored in Alchemist. After transmitting the data in an RDD to Alchemist, Alchemist returns an {\tt AlMatrix} object, which contains a unique ID identifying the matrix to Alchemist, as well as other information such as the dimensions of the matrix.

Similarly, for every output matrix that an MPI-based routine creates, an {\tt AlMatrix} object is returned to the application. These {\tt AlMatrix} objects allow the user to pass the distributed matrices within Alchemist from one library function to the next. Only when the user explicitly converts this object into an RDD will the data in the matrix be sent from Alchemist to Spark. 

In this example, two {\tt AlMatrix} objects get returned ({\tt AlQ} and {\tt AlR}), and they get converted to {\tt IndexedRowmatrix} objects using the {\tt AlMatrix} class's {\tt toIndexedRowmatrix()} function. Only at this point does the data get transmitted from Alchemist to Spark.
\item The application has to stop the {\tt AlchemistContext} similarly to how an instance of {\tt SparkContext} is stopped. 
\end{itemize}
Note that the API may be tweaked in future releases.

\section{Experiments}
\label{sec:experiments}

All of the following experiments were run on the NERSC supercomputer Cori, a Cray XC40. In particular, we used the Cori Phase 1 system (also known as the Cori Data Partition system), which has two 2.3 GHz 16-core Intel Haswell processors on each of its 2,388 nodes, and 128 GB of memory per node.

\subsection{CG Solver for Speech Classification}
\label{subsec:cgkrr}

Solving large systems of equations is an important and common task in many areas of research. Our primary focus here is on the solution of a large linear system arising in the area of speech classification.

The data set is a modified version of the TIMIT speech corpus, which was designed to provide speech data for acoustic-phonetic studies and for the development and evaluation of automatic speech recognition systems~\cite{TIMIT1993}. TIMIT was commissioned by DARPA and contains broadband recordings of 630 male and female speakers of eight major dialects of American English, each reading ten phonetically rich sentences. The TIMIT corpus includes hand-verified, time-aligned orthographic, phonetic and word transcriptions as well as a 16-bit, 16kHz speech waveform file for each utterance. 

A preprocessing pipeline was introduced in~\cite{Huang2014} that provided a modified version of TIMIT with $2,251,569$ training examples, $100,000$ test examples, and $440$ raw features. This preprocessing pipeline has subsequently been used elsewhere, for instance~\cite{Tu2016}, and we also use it here. 

In this classification problem, a given feature vector is assigned into one of 147 different classes. Consequently, for each feature vector $x$ used for training, there is an associated $147$-dimensional label vector $y$ that has only one non-zero entry. If $x$ is in the $i$-th class, then we hope to find a weighting matrix $W$ such that the $147$-dimensional vector $x\,W$ is all-zero except for the $i$-th entry. We let $X$ denote the feature-space data matrix containing all the training examples, and $Y$ the associated label matrix. 

It turns out that the original data set is not particularly expressive since there are just $440$ features, leading \cite{Huang2014} to use the random feature method proposed in~\cite{Rahimi2008} to increase the expressiveness.
This approach, however, has the drawback that the minimum random feature space dimensionality needed for competitive performance tends to be very large when the number of training examples and/or the dimension of original features are large, and this poses a scalability problem. 
In this study we use random feature space dimensions of an order of magnitude of $10,000$, and thereby we expand the feature matrix $X$, which would otherwise be $2,251,569 \times 440$, into a larger feature matrix, say $2,251,569 \times 10,000$. This is still an over-determined system, but it is much more expressive, and the training and test errors are lower.

In the simulations involving Alchemist, we send the original $2,251,569 \times 440$ feature matrix from Spark to Alchemist, not an expanded one. The feature matrix is instead expanded within Alchemist using the random feature method. This is done to ensure that the size of the matrix that we are sending is consistent from one simulation to the next, and it is also significantly cheaper to do the expansion within Alchemist rather than transferring a feature matrix that is several TB in size.

Recall that we want to find the weighting matrix $W$. We do so by solving the regularized minimization problem
$$
	\argmin_W \dfrac{1}{n}\,||X\, W - Y ||_F^2 + \lambda ||W||_F^2,
$$
where $||\,\cdot\, ||_F$ denotes the Frobenius norm, $n$ is the number of training examples, and $\lambda$ is a regularization parameter. This minimization problem is equivalent to solving the linear system
$$
	(X^T X + n\lambda I)\,W = X^T Y,
$$
and because this system is symmetric positive definite, we can use the \textit{conjugate gradient} (CG) method to solve it~\cite{Saad2003}. As is usual for the solution of large linear systems, the CG method is an iterative method.

\begin{table}
\centering
\begin{tabular}{ c | c | c | c | c }
    \hline
	data set & Matrix & Matrix Dimensions & Spark & Alchemist \\
    \hline
    Original & Feature & $2,251,569 \times 440$ &   &   \\
             & Label & $2,251,569 \times 147$ &   &   \\
    \hline
    Modified & Feature & $2,251,569 \times 10,000$ & Yes & Yes \\
             & Feature & $2,251,569 \times 20,000$ & No & Yes \\
             & Feature & $2,251,569 \times 30,000$ & No & Yes \\
             & Feature & $2,251,569 \times 40,000$ & No & Yes \\
             & Feature & $2,251,569 \times 50,000$ & No & Yes \\
             & Feature & $2,251,569 \times 60,000$ & No & Yes \\
    \hline
\end{tabular}
\vspace{1em}
  \caption{Summary of matrices used in the speech classification study. We were unable to run simulations with Spark using more than 10,000 features.}
  \label{table:timit_summary}
\end{table}


We use a modification of the MPI-based implementation of CG found in the libSkylark library~\cite{Skylark2016}. Skylark is an open source software library for distributed randomized numerical linear algebra, with applications to machine learning and statistical data analysis; it is built atop the Elemental library, making it ideally suited for use with the current implementation of Alchemist. We wrote our own version of CG in Spark, since no suitable implementations were available in MLlib.

We were unable to get Spark to perform the conjugate gradient method on feature matrices that contained more than 10,000 features, but having a large number of features did not prove to be a problem for Alchemist. Table~\ref{table:timit_summary} summarizes the sizes of the feature matrices that were used in  this study (observe that the ``smaller'' dimension is gradually increasing), and with which system they were used.

\begin{table}
\centering
\begin{tabular}{ c || c | c || c }
    \hline
	Number & \multirow{2}{*}{System} & Iteration time cost & \textbf{Computation} \\
    of nodes & & in s (mean $\pm$ s.d.) & \textbf{time (in s)} \\
    \hline
    \multirow{2}{*}{20} & Spark & $75.3 \pm 18.8$ & \textbf{39,743} \\
     & Alchemist & $2.5 \pm 0.4$ & \textbf{1,330} \\
    \hline
    \multirow{2}{*}{30} & Spark & $55.9 \pm 8.7$ & \textbf{29,443} \\
     & Alchemist & $1.5 \pm 0.1$ & \textbf{789} \\
    \hline
    \multirow{2}{*}{40} & Spark & $40.6 \pm 9.1$ & \textbf{21,442} \\
     & Alchemist & $1.2 \pm 0.2$ & \textbf{646} \\
    \hline
\end{tabular}
\vspace{1em}
  \caption{Per-iteration costs of Spark vs Alchemist for the CG method on the $2,251,569 \times 10,000$ problem. Alchemist uses a modified version of the CG method implemented in the libSkylark library.}
  \label{table:cg_spark_vs_alchemist}
\end{table}

\begin{table}
\centering
\begin{tabular}{ c | c | c | c | c }
   \cline{1-5}
    \multicolumn{1}{c|}{Number} & \multicolumn{1}{|c|}{Data set} & \multicolumn{3}{|c}{Feature matrix transfer times (in s)} \\
    \cline{3-5}
   \multicolumn{1}{c|}{of Spark} & \multicolumn{1}{|c|}{creation} & \multicolumn{3}{|c}{Number of Alchemist processes} \\
   \multicolumn{1}{c|}{processes} & \multicolumn{1}{|c|}{times (in s)} & \quad 20 \quad & \quad 30 \quad & \quad 40 \quad \\
    \hline
    \multicolumn{1}{c|}{2}  & 777.0 & 580.1 & 874.9 & 1,021.6 \\
    \multicolumn{1}{c|}{10} & 238.2 & 166.4 & 198.0 & 222.9 \\
    \multicolumn{1}{c|}{20} & 217.5 & 149.5 & 165.7 & 185.4 \\
    \multicolumn{1}{c|}{30} & 228.0 & 163.1 & 157.6 & * \\
    \multicolumn{1}{c|}{40} & 209.4 & 312.4 & * & * \\
    \hline
    \multicolumn{2}{c|}{Computation time} & \quad 1,330.08 \quad  & \quad 788.51 \quad & 646.6 \\
    \hline
  \end{tabular}
\vspace{1em}
  \caption{Time required to transfer the feature matrix from Spark to Alchemist, for different node allocation configurations, with a maximum of 60 nodes available. For comparison purposes, we have also included the time it takes to perform the CG computations. The reported transfer times are the average of three separate runs. }
  \label{table:transfer_times}
\end{table}

See Table~\ref{table:cg_spark_vs_alchemist} for a comparison of the per-iteration and total computation time costs between Spark and Alchemist for the case of 10,000 features. We see an order of magnitude improvement in overall running time of the CG computations. This is not surprising, since the CG method is an iterative method and we therefore expect Spark to have high overheads, which leads to the significantly higher per-iteration time cost. In contrast, the main overhead when using Alchemist is the time it takes to transfer the input data from the Spark processes to the Alchemist processes, and (depending on the application) to transfer the output from the Alchemist processes to the Spark processes. In this experiment we do not need to worry about the time it takes to transfer the output data, but transferring the input data is non-negligible.

The time taken for the data transfer depends on several factors. Most significantly, of course, is the size of the matrix, but also the layout of the data in the RDD compared to the desired layout in the distributed matrix on the library side, the physical proximity of the nodes running the Spark processes to the nodes running the Alchemist processes (this is less important on supercomputers with fast interconnects), network load and reliability, as well as the number of Spark processes relative to the number of Alchemist processes.

We investigate the latter here by looking at the transfer times of the feature matrix for different numbers of Spark and Alchemist processes. We hold the number of features fixed at 10,000. The results can be seen in Table~\ref{table:transfer_times}. Notice that the transmission time decreases as the number of Spark executors increases, which is not surprising because in this way more data can be sent to Alchemist concurrently. It is also interesting that, based on the observations for 20 Alchemist nodes, it appears that the transmission time is minimized when using the same number of Spark and Alchemist processes, although a more thorough investigation needs to be performed to see if this holds true in general.

Also, the cost of sending the feature matrix from Spark to Alchemist will increase relative to the total computation time as the number of Alchemist processes increases. This is simply a side effect of having a reduced CG computation time when using more Alchemist processes. The times reported here are the average of three different runs, since there is some variability in the transfer times due to factors such as network load; see~\cite{Gittens2018b} for an investigation of transfer times of a randomly-generated, dense 400GB matrix for two different matrix shapes.

\begin{table}
\centering
\begin{tabular}{ c || c || c }
    \hline
	Number of  & Iteration time cost & \textbf{Computation} \\
    features & in ms (mean $\pm$ s.d.) & \textbf{time (in s)} \\
    \hline
    10,000 & $1,490.6 \pm 65.7$ & \textbf{788.5} \\
    20,000 & $2,895.8 \pm 39.8$ & \textbf{1,534.8} \\
    30,000 & $4,317.0 \pm 48.4$ & \textbf{2,270.7} \\
    40,000 & $5,890.4 \pm 67.8$ & \textbf{3,104.2} \\
    50,000 & $7,286.9 \pm 26.1$ & \textbf{3,854.8} \\
    60,000 & $8,794.9 \pm 99.0$ & \textbf{4,643.7} \\
    \hline
\end{tabular}
\vspace{1em}
  \caption{Per-iteration and total time costs for the CG computations using Alchemist for different numbers of features. Alchemist ran on 30 nodes, and the time to transmit the data from Spark (also running on 30 nodes) was 169.6s on average. We see that the data transmission overhead becomes less significant as the number of features grows.}
  \label{table:alchemist_run_times_features}
\end{table}

Finally, in Table~\ref{table:alchemist_run_times_features} we show the per-iteration and total computation times for the CG computations using Alchemist for an increasing number of features. The per-iteration cost increases linearly with the number of features, and performing these computations with Alchemist therefore appears to be scalable, although a more thorough investigation is required to see how large the number of features can get before this behavior breaks down.

For all of these computations we have set $\lambda = 10^{-5}$. This parameter affects the condition number of the linear system and thereby the convergence, i.e., the number of iterations to reach a fixed precision; it does not affect the per-iteration elapsed time. When using $10,000$ random features and $\lambda = 10^{-5}$, CG takes approximately $526$ iterations to reach a relative residual of machine precision.
\subsection{Truncated SVD of Large Data Sets}
\label{subsec:truncated_svd}

An important procedure in the analysis of large data sets, especially in scientific research, is principal component analysis (PCA), which is used to reduce the dimensionality of large data sets and to uncover meaningful structure. PCA relies on the singular value decomposition (SVD), so to demonstrate the advantage of using Spark+Alchemist over just Spark, a study was performed in~\cite{Gittens2018b} to compare the times it takes to perform the decomposition. 

In that study, the rank-$k$ truncated SVD (meaning we compute only the first $k$ singular values, for some modest value of $k$) of randomly generated double-precision dense matrices, ranging in size from 25GB to 400GB, is computed using the MLlib implementation in Spark vs. a custom MPI-based implementation in Alchemist\footnote{Both implementations make use of ARPACK~\cite{ARPACK1997} to compute the eigenvalues of the Gram matrix.}. There is a substantial improvement in total execution times when using Spark+Alchemist vs. just Spark, and it clearly underlines the fact that Alchemist can significantly improve the workflow when analyzing large data sets, and that this is done in a scalable manner. 

In this section we expand on these results and use a real data set of ocean temperature data. Climate scientists use the truncated SVD to extract important patterns from the measurement of climatic variables, but traditionally their analyses have been limited to that of two-dimensional fields. We applied Alchemist's truncated SVD to the analysis of a three-dimensional field covering an approximately 67-month subset of the Climate Forecast System Reanalysis (CFSR) data set~\cite{cfsr}. This data set is distributed by the National Climatic Data Center and contains measurements of various atmospheric, oceanic, and land weather quantities of interest to meteorologists and climate scientists, sampled at a six-hourly time resolution on a grid of $0.5\deg$ latitude by $0.5\deg$ longitude resolution at varying altitudes and depths. In particular, we consider the global ocean temperature readings between January 1979 and mid-1984 at 40 subsurface levels, yielding a 400GB double precision matrix of size $6,177,583 \times 8,096$. 

\begin{table}
\centering
\begin{tabular}{ c | c || c | c | c | c || c }
    \hline
	\multicolumn{2}{ c||}{Number} & & S $\Rightarrow$ A & SVD & S $\Leftarrow$ A &  \\
    \multicolumn{2}{ c||}{of nodes} & Data load & data transfer & compute & data transfer & \textbf{Total run} \\
    \cline{1-2}
    \,\,S\,\, & A & time (in s)  & time (in s) & time (in s) & time (in s) & \textbf{time (in s)} \\
    \hline
    12 & 0 & 38.0 & NA & 553.1 & NA & \textbf{553.1} \\
    10 & 12 & 38.0 & 62.5 & 48.6 & 10.8 & \textbf{121.9} \\
    2 & 12 & 27.23 & NA & 48.6 & 21.1 & \textbf{69.7} \\
    \hline
\end{tabular}
\vspace{1em}
  \caption{Run-times for three use cases of Spark (S) and Alchemist (A) for computing the rank-20 SVD of the abridged ocean temperature data set: 
 (1) Spark loads the data and computes the SVD, 
 (2) Spark loads the data and Alchemist computes the SVD,
 (3) Alchemist both loads and computes the SVD before transferring the result back to Spark. In this example, using Alchemist leads to speedups of about 4.5x and 7.9x compared to only using Spark. The total run times do not include the time it takes to load the data. }
  \label{table:svd_spark_vs_alchemist_run_times}
\end{table}

We consider three use cases: one where Spark is used to both load and decompose the data set, one where Spark is used to load the data set and Alchemist is used to compute the truncated SVD, and one where Alchemist is used to both load and decompose the data set before passing the results back to Spark. In all three use cases, we allocate 12 nodes to the system computing the rank-$20$ truncated SVD, whether Spark or Alchemist, to compare their ability to use the same resources. In the second use case, we use 10 nodes of Spark to load and transfer the data set to Alchemist, and in the final use case, we use just two nodes of Spark to receive the decomposition from Alchemist. The data set is supplied in both HDF5 format, which Alchemist can read directly, and the Spark-readable Parquet~format.

Table~\ref{table:svd_spark_vs_alchemist_run_times} compares the run-times of these three setups when computing the truncated SVD of the ocean temperature data set. We see that offloading the computation to Alchemist results in a significant drop in the total runtime. Reading the data set using Spark and letting Alchemist do the computations leads to a speedup of about 4.5x compared to just using Spark. A speedup of about 7.9x is achieved by loading and processing the data set in Alchemist and sending the results to Spark.

This experiment demonstrates that offloading linear-algebraic tasks from Spark onto Alchemist can result in significant decreases in runtime, even when the overheads of the data transfer are taken into account.

\begin{figure}[htp]
\centering
\includegraphics[width=\textwidth]{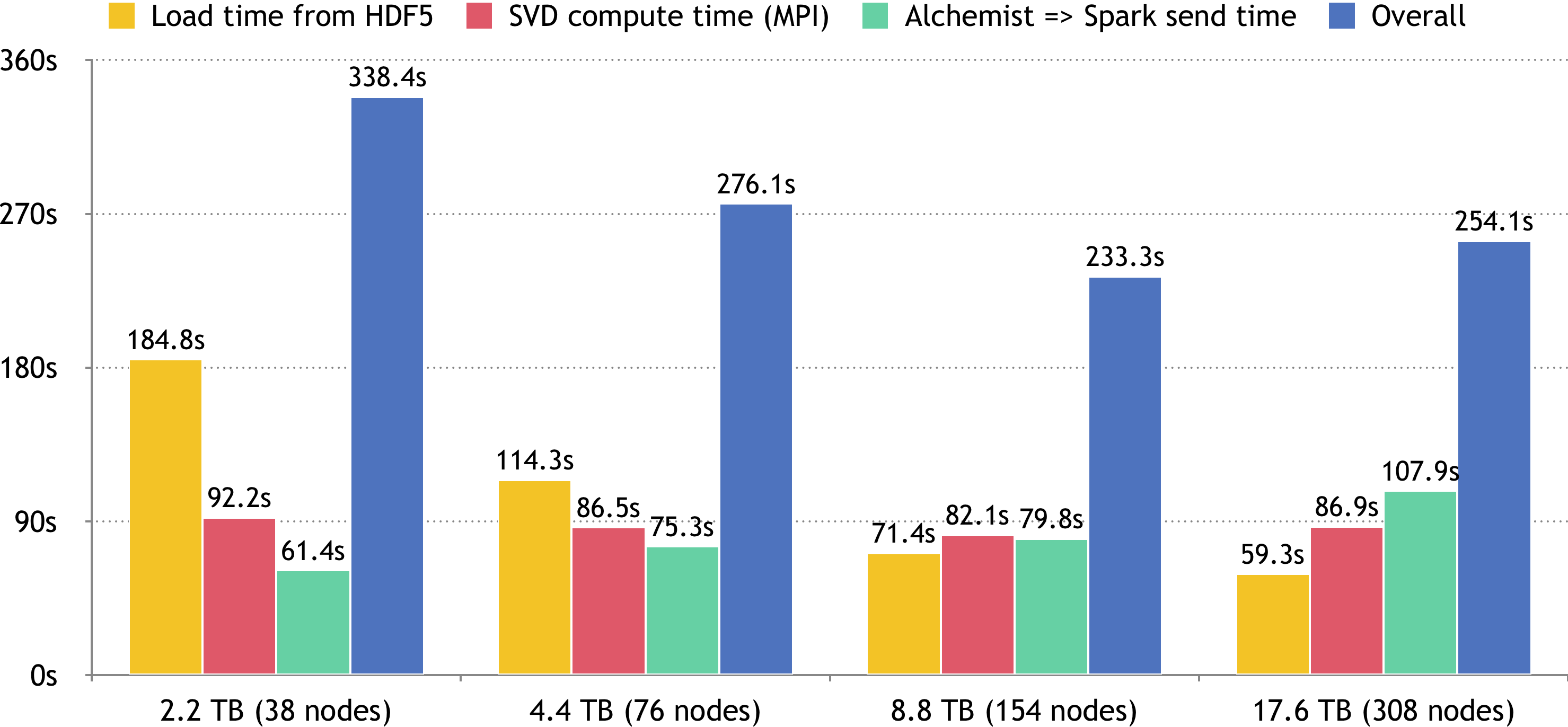}
\caption{Timing data for SVD of a 2.2TB data set loaded from file and replicated column-wise.}
\label{fig:SVD_replicated}
\end{figure}

The 400GB ocean data set used above is actually an abridged version of a 2.2TB ocean temperature data set with dimensions $6,177,583 \times 46,752$. We use this larger data set to illustrate that using Alchemist leads to computations that are scalable even when dealing with data in the order of 10TB to 20TB. This is done by loading the data from the HDF5 file directly in Alchemist and replicating it column-wise a certain number of times to attain data sets of sizes 2.2TB, 4.4TB, 8.8TB and 17.6TB. The rank-$20$ truncated SVD is performed on the data, and the singular vectors and values are then sent from Alchemist to Spark, where they could then be used for further analysis. Figure~\ref{fig:SVD_replicated} shows the times it takes these different operations, along with the number of nodes that were used in the computations.

Note the (weak) scaling of the SVD computations. As the number of processes and the size of the data set get doubled, the time it takes to perform the SVD computation is consistent. Unsurprisingly, the total load time from HDF5 decreases as the number of nodes increases, and the time to send the results to Spark increases as the size of the data set increases. In all cases we used one Spark executor to store the results.

\section{Discussion and Conclusion}
\label{sec:conclusion}

Apache Spark has made a significant impact in recent years and provides a valuable platform for the analysis of large data sets. However, due to large overheads arising within the Spark framework, there are some types of computations, in particular linear algebra routines, that have been shown to be significantly slower than equivalent computations performed using MPI-based implementations.

Alchemist addresses this by interfacing between Spark and MPI-based libraries, and we have shown how it can significantly speed up important linear algebra routines such as the SVD.
Among other things, this enables the scaling of low-rank PCA computation to data sets of size 10TB to 20TB.
This improved performance comes with comparatively little overhead, essentially just the cost of moving the data over the network between Alchemist and Spark nodes, while still retaining the benefits of working in the Spark environment. This makes the use of Spark more attractive to potential users who want the combination of the performance of MPI with the simple interface and extensive ecosystem of Spark. 

\subsection{Limitations and Constraints}

Although there are various benefits that Alchemist would provide to a data analysis workflow that makes heavy use of linear algebra routines, there are some limitations and constraints in using the system:
\begin{itemize}
\item Two copies of the data are required: the RDD used by the Spark application, and the same data stored as a distributed matrix by Alchemist. This is a necessary limitation, since MPI-based codes are not able to access the data in the RDD directly. 
\item Alchemist will also require additional storage at runtime for storing output from the computations, and temporary storage used by the MPI-based routines during the computations. Due to the inelastic nature of MPI, Alchemist must run on a sufficient number of nodes before starting the computations.
\item Overhead is incurred when transferring data between Spark and the MPI libraries over a network, and is subject to network disruptions and overload. However, for computationally-intense operations the cost associated with this data transfer is negligible when compared to the overhead that would have been incurred by Spark, and it is less than when using the other methodologies for transferring data that were mentioned in Section \ref{subsec:transmitting_data}.
\item While we have looked at linear algebra routines in this paper, MPI-based libraries for other applications can also be used with Alchemist, which would be particularly interesting if machine learning libraries are available. Unfortunately, it appears that there are few MPI-based machine learning projects under development at the moment, and they tend to be quite limited in their scope. One exception to this is MaTEx~\cite{Matex2017}.
\end{itemize}

\subsection{Future Work}

Further improvements and refinements to the system are ongoing and we summarize some of these here:
\begin{itemize}
\item A Python interface is being implemented that allows users of PySpark to use Alchemist. This would also allow users of other Python development environments, for instance Jupyter Notebooks, to access MPI-based libraries to help analyze distributed data sets. 
\item Development of Alchemist has thus far taken place on Cori, a NERSC supercomputer. To enable Alchemist to run on clusters and the cloud, it is most convenient to run it from inside some container such as a Docker image.
\item Although Alchemist at present only directly supports MPI-based libraries that make use of Elemental, it is nonetheless possible to use MPI-based libraries built on top of other distributed linear algebra packages, for instance ScaLAPACK and PLAPACK. However, this requires the use of wrapper functions to convert Elemental's distributed matrices to the appropriate format, which will incur additional overhead, possibly including an additional copy of the data in memory. Support for some other distributed linear algebra libraries will be added in the future.
\end{itemize}

\section*{Acknowledgements}
This work was partially supported by NSF, DARPA, and Cray Inc.

\bibliographystyle{plain}
\bibliography{Bibliography} 

\end{document}